# Quantum Algorithms for Modulated Circulant Matrix–Vector Multiplication

Kimy Agudelo

*Departamento de Física, Facultad de Ciencias. Universidad Católica del Norte. Av. Angamos 0610 Antofagasta, Chile*

Aldo Quelopana

*Departamento de Ingeniería de Sistemas y Computación, Facultad de Ingeniería y Ciencias Geológicas. Universidad Católica del Norte. Av. Angamos 0610 Antofagasta, Chile*

Cristina Manzaneda*

*Departamento de Matemáticas, Facultad de Ciencias. Universidad Católica del Norte. Av. Angamos 0610 Antofagasta, Chile*

**Abstract**

Modulated circulant matrices form a special class of $N-$parametric circulant matrices, recently introduced in the literature, with a structured spectral decomposition based on a Vandermonde-type basis. Motivated by this definition, in this work we define the *Modulated Quantum Fourier Transform* (MQFT), a quantum primitive tailored to this matrix family. We show that, under an oracle model for the modulation parameters, the MQFT can be implemented with asymptotic gate complexity $O(\log^2 N)$, matching that of the standard quantum Fourier transform. Using this transform, we develop two quantum algorithms for modulated circulant matrix–vector multiplication, derive their complexity, and provide explicit quantum circuit constructions. Additionally, we present numerical simulations that validate the proposed framework, achieving output fidelities above 0.97 in the non-padded cases and showing close agreement between theoretical and measured success prob-

*Corresponding author

*Email addresses:* `kimy.agudelo@alumnos.ucn.cl` (Kimy Agudelo), `aldo.quelopana@ucn.cl` (Aldo Quelopana), `cmanzaneda@ucn.cl` (Cristina Manzaneda )

abilities.

## 1. Introduction

Quantum computing has attracted significant attention due to its potential to accelerate the solution of certain computational problems beyond what is achievable with classical methods. A prominent example in quantum linear algebra is the Harrow–Hassidim–Lloyd (HHL) algorithm [1], which provides an exponential speedup for estimating specific properties of the solution of linear systems under well-defined assumptions on sparsity, conditioning, and data access. At a high level, the HHL algorithm combines quantum phase estimation with eigenvalue-dependent controlled rotations to encode the inverse of a matrix in its spectral decomposition, followed by an uncomputation step that yields a quantum state proportional to the desired solution.

A key feature of the HHL framework is that its computational complexity depends not only on the dimension of the matrix but also on its condition number, which quantifies the sensitivity of the solution to perturbations and is an intrinsic property of the problem. As a consequence, directly constraining the condition number is often impractical. This observation has motivated growing interest in identifying classes of structured matrices whose algebraic and spectral properties can be exploited to design more efficient quantum linear algebra procedures.

Beyond the original scope of the HHL algorithm, structured matrices have therefore received particular attention. Among them, circulant matrices constitute a notable example, as they admit a simple and well-understood spectral decomposition given by the discrete Fourier transform. This property enables efficient diagonalization and has long been exploited in classical algorithms to accelerate linear algebraic .

Circulant matrices have been extensively studied in both classical and quantum settings. In the classical case, the product of an $N \times N$ circulant matrix and a vector can be computed in $O(N \log N)$ time using the Fast Fourier Transform (FFT) [2]. Moreover, by embedding a Toeplitz matrix into a circulant matrix, Golub and Van Loan showed that Toeplitz matrix–vector multiplication can also be performed in $O(N \log N)$ time [2]. In the quantum setting, several representations and algorithmic frameworks for cir-culant matrices have been proposed, enabling their efficient manipulation within quantum circuits [3, 4, 5]

In this work, we focus on a generalization of the circulant structure, namely the so-called *$N$-parametric circulant matrices*, as introduced and analyzed in [6]. These matrices extend the classical circulant framework by allowing additional degrees of freedom in their parameterization, while preserving a structured spectral form amenable to quantum processing. We propose quantum algorithms for matrix–vector multiplication with $N$-parametric circulant matrices that exploit their spectral structure through quantum phase estimation and eigenvalue-dependent transformations. Under standard assumptions on data access and conditioning, the proposed algorithms achieve improved computational complexity when compared to classical approaches.

Let $n \geq 0$ be an integer, and $N = 2^n$. The *quantum Fourier transform* on $n$ qubits is defined by

$$F_N : |x\rangle \longrightarrow \frac{1}{\sqrt{2^n}} \sum_{y=0}^{2^n-1} e^{2\pi i x y/2^n} |y\rangle, \quad (0 \leq x \leq 2^n - 1). \tag{1}$$

The complexity of the quantum Fourier transform is $O(n^2)$.

**Notation:** In what follows in manuscript $N$ denotes the order of the matrix and $n$ denote the number of qubits. The field of complex numbers is denoted as usual by $\mathbb{C}$. The identity matrix $N \times N$ is denoted by $I_N$, and $\operatorname{diag}(d_1, \ldots, d_N)$ represents the diagonal matrix with diagonal entries $d_1, d_2, \ldots, d_n$. For any square matrix $A$, invertible, $A^{-1}$ is its inverse, $A^T$ is its transpose, and $A^\dagger$ is its conjugate transpose. We denote by $\mathbf{e}_i$ the $i$-th column of $I_N$. The Kronecker product of two matrices $A$ and $B$ is denoted by $A \otimes B$. The null matrix $N \times N$ is denoted by $0_N$. Furthermore, $F = \frac{1}{\sqrt{N}} \left( \omega^{(i-1)(j-1)} \right)_{ij}$, $1 \leq i, j \leq N$, is the Discrete Fourier Transform (DFT), where $\omega = \exp(\frac{2\pi \mathrm{i}}{N})$ is the primitive root $n$ -th of unity, with $\mathrm{i} = \sqrt{-1}$. The vector $\mathbf{1}_N \in \mathbb{R}^N$ is the vector $(1, 1, \ldots, 1)$.

## 2. The modulated circulant matrix

In [6] a new generalization of the circulating structure is defined. Furthermore, it is shown that this generalization has similar properties and spectral decomposition of the circulating structure. In this section, we focus on an important particular instance of the $N-$parametric matrices, which they will call *the modulated circulant matrix*. We introduce the corresponding matrix and establish several algebraic properties that will play a crucial role in the subsequent section.

**Definition 1.** *Let $N$ non-negative integer number and $a_0,\ldots,a_{N-1}$ be non-zero complex numbers with $|a_j|=1$, for all $j=0,1,\ldots,N-1$. The modulated circulant matrix, denoted by $M_{\bar{a}}(\boldsymbol{x})$ is the following matrix:*

$$M_{\bar{a}}(\boldsymbol{x})=\sum_{r=0}^{N-1}x_r(T_{\bar{a}})^r, \tag{2}$$

*where $\boldsymbol{x}=(x_0,\ldots,x_{N-1})\in\mathbb{C}^N$ and*

$$T_{\bar{a}}=\begin{bmatrix} 0 & a_0 & 0 & \cdots & 0 \\ 0 & 0 & a_1 & \ddots & \vdots \\ \vdots & \ddots & \ddots & \ddots & a_{N-2} \\ a_{N-1} & 0 & \cdots & 0 & 0 \end{bmatrix}, \tag{3}$$

*$\bar{a}=(a_0,a_1,\ldots,a_{N-1})$ and $a_i$, $i=0,\ldots,N-1$, are called the parameters of the matrix.*

**Remark 1.** *We repair that $T_{\bar{a}}$ is a unitary matrix. In addition, is in particular a modulated circulant matrix, this is $T_{\bar{a}}=M_{\bar{a}}(0,1,0,\ldots,0)$. Furthermore, in the particular cases $\bar{a}=\boldsymbol{1}_N$ and $\bar{a}=(\boldsymbol{1}_{N-1},k)$, the matrix $M_{\bar{a}}(\boldsymbol{x})$ reduces to a circulant matrix and a k-circulant matrix, respectively.*

**Definition 2.** *Let $N\geq 2$ and $a_i\in\mathbb{C}-\{0\}$, for all $i$. Consider $\gamma$ any $N$-th root of $\prod_{j=0}^{N-1}a_j$, we define the Modulated Discrete Fourier Transform (MDFT) by*

$$(F_{\bar{a}})_{ij}=\frac{1}{\sqrt{N}}\gamma_i\omega^{(i-1)(j-1)};\quad 1\leq i,j\leq N, \tag{4}$$

*where $\gamma_j=\gamma_{j-1}\frac{\gamma}{a_{j-1}}$, $j=1,\ldots,N-1$ and $\gamma_0=1$.*

**Remark 2.** *The $F_{\bar{a}}^{\dagger}$ is given by*

$$(F_{\bar{a}}^{\dagger})_{ij}=\frac{1}{\sqrt{N}}\zeta_i\omega^{-(i-1)(j-1)};\quad 1\leq i,j\leq N, \tag{5}$$

*where* $\zeta_j = \zeta_{j-1}\dfrac{a_{j-1}}{\gamma}$, $j = 1, \ldots, N-1$ *and* $\zeta_0 = 1$.

**Example 1.** *Let* $\bar{a} = (1, e^{i\pi/4}, -1, e^{-i\pi/4})$, *for* $\gamma = e^{i\pi/4}$, *according to the formula* (4) *and* (5), *respectively, we have*

$$F_{\bar{a}} = \frac{1}{2}\begin{pmatrix} 1 & 1 & 1 & 1 \\ e^{i\pi/4} & -e^{i3\pi/4} & -e^{i\pi/4} & e^{i3\pi/4} \\ e^{i\pi/4} & -e^{i\pi/4} & e^{i\pi/4} & -e^{i\pi/4} \\ -e^{i\pi/2} & -e^{i\pi} & e^{i\pi/2} & e^{i\pi} \end{pmatrix}$$

*and*

$$F^{\dagger}_{\bar{a}} = \frac{1}{2}\begin{pmatrix} 1 & e^{-i\pi/4} & e^{-i\pi/4} & -e^{-i\pi/2} \\ 1 & -e^{-i3\pi/4} & -e^{-i\pi/4} & -e^{-i\pi} \\ 1 & -e^{-i\pi/4} & e^{-i\pi/4} & e^{-i\pi/2} \\ 1 & e^{-i3\pi/4} & -e^{-i\pi/4} & e^{-i\pi} \end{pmatrix}$$

*Furthermore,*

$$T_{\bar{a}} = \begin{pmatrix} 0 & 1 & 0 & 0 \\ 0 & 0 & e^{i\pi/4} & 0 \\ 0 & 0 & 0 & -1 \\ e^{-i\pi/4} & 0 & 0 & 0 \end{pmatrix}$$

*and*

$$M_{\bar{a}}(x_0, x_1, x_2, x_3) = \begin{pmatrix} x_0 & x_1 & e^{i\pi/4}x_2 & -e^{i\pi/4}x_3 \\ -x_3 & x_0 & e^{i\pi/4}x_1 & -e^{i\pi/4}x_2 \\ -e^{-i\pi/4}x_2 & -e^{-i\pi/4}x_3 & x_0 & -x_1 \\ e^{-i\pi/4}x_1 & e^{-i\pi/4}x_2 & x_3 & x_0 \end{pmatrix} \tag{6}$$

*Note that this matrix is neither a Toeplitz matrix nor a k-circulant matrix; it is a special case of* $n-$*parametric circulant matrix. Also, is diagonalizable by the unitary matrix* $F_{\bar{a}}$ *and generated by* $T_{\bar{a}}$.

**Theorem 1.** *A spectral decomposition of the* $T_{\bar{a}}$ *matrix is given by*

$$T_{\bar{a}} = F_{\bar{a}}\Lambda_{\gamma}F^{\dagger}_{\bar{a}}, \tag{7}$$

*where* $\Lambda_{\gamma} = \mathrm{diag}\left(\gamma, \gamma\omega, \ldots, \gamma\omega^{N-1}\right)$ *and* $\gamma$ *is any* $N$*-th root of* $\prod_{j=0}^{N-1} a_j$.

*Proof.* Note that, for $j = 1, \dots, N$, $T_{\bar{a}}\mathbf{v}_j = \gamma\omega^{j-1}\mathbf{v}_j$ where

$$\mathbf{v}_j = \frac{1}{\sqrt{N}} \begin{bmatrix} 1 & \frac{\gamma\omega^{j-1}}{a_0} & \frac{(\gamma\omega^{j-1})^2}{a_0 a_1} & \cdots & \frac{(\gamma\omega^{j-1})^{N-1}}{a_0 a_1 \cdots a_{N-2}} \end{bmatrix}^T,$$

is the $j$-th column of $F_{\bar{a}}$. □

**Corollary 1.** *A spectral decomposition of the $M_{\bar{a}}$ matrix is given by*

$$\operatorname{diag}\{F_{\bar{a}}^{\dagger}\boldsymbol{x}\} = F_{\bar{a}} M_{\bar{a}} F_{\bar{a}}^{\dagger}, \tag{8}$$

*Proof.* Note that, if $T_{\bar{a}}\mathbf{v}_j = \lambda_j \mathbf{v}_j$ then

$$M_{\bar{a}}\mathbf{v}_j = \sum_{r=0}^{N-1} c_r \lambda_j^r \mathbf{v}_j,$$

Therefore, the result is an immediate consequence of Theorem 1 and Definition 2. □

The following results are a direct consequence of [6, Theorem 1, Theorem 2], considering $s = 1$ and $|a_j| = 1$, for all $j = 0, 1, \dots, N-1$.

**Theorem 2.** *A square matrix $A$ is an $N-$parametric circulant matrix if and only if*

$$A = F_{\bar{a}} C F_{\bar{a}}^{\dagger},$$

*for some circulant matrix $C$ and some $\bar{a} \in \mathbb{C}^N$.*

**Corollary 2.** *Let $M_{\bar{a}}$ be a modulated circulant matrix with first row $x_0, x_1, \dots, x_{N-1} \in \mathbb{C}$ and $\gamma$ any $N$-th root of $\prod_{j=0}^{N-1} a_j$. Then*

$$\mu_k = \sum_{j=0}^{N-1} x_j (\gamma\omega^k)^j, \quad k = 0, 1, \dots, N-1, \tag{9}$$

*are the eigenvalues of $M_{\bar{a}}(x)$. Moreover $M_{\bar{a}} = M_{\bar{a}}(x_0, x_1, \dots, x_{N-1})$, where*

$$x_k = \frac{1}{N} \sum_{j=0}^{N-1} (F_{\bar{a}}^{\dagger} x)_j (\gamma\omega^j)^{-k}, \quad k = 0, 1, \dots, N-1. \tag{10}$$

**Example 2.** *Applying the formulas from the preceding corollary to matrix* (6) *in Example 1 yields the following eigenvalues:*

$$\begin{aligned}
\mu_0 &= x_0 + e^{i\pi/4}x_1 + e^{i\pi/2}x_2 + e^{i3\pi/4}x_3\\
\mu_1 &= x_0 - ie^{i\pi/4}x_1 - e^{i\pi/2}x_2 + ie^{i3\pi/4}x_3\\
\mu_2 &= x_0 - e^{i\pi/4}x_1 + e^{i\pi/2}x_2 - e^{i3\pi/4}x_3\\
\mu_3 &= x_0 + ie^{i\pi/4}x_1 - e^{i\pi/2}x_2 - ie^{i3\pi/4}x_3
\end{aligned}$$

*If the values $\mu_j$ were known, we could obtain $M_{\bar{a}}(x_0, x_1, x_2, x_3)$ by applying the formula in* (10).

**Corollary 3.** *If $A_{\bar{a}}$ and $B_{\bar{a}}$ are modulated circulant matrices of order $N$ and $\alpha$ and $\beta$ are scalars, then $(A_{\bar{a}})^{-1}$ $(A_{\bar{a}})^T$, $(A_{\bar{a}})^\dagger$, $\alpha A_{\bar{a}} + \beta B_{\bar{a}}$ and $A_{\bar{a}}B_{\bar{a}}$ are modulated circulant matrices.*

We note that, as a consequence of the Corollary above, the set of modulated circulant matrices forms a subfield in the matrix space. On the other hand, an important characterization of the circulant structure is that a square matrix $A$ is a circulant matrix if and only if it commutes with the matrix $T$ defined in (3) with $\bar{a} = \mathbf{1}_N$, that is, $AT = TA$. In this sense, we have a more general result.

**Theorem 3.** *Let $A = [a_{i,j}]$ be an $N \times N$ matrix. Then $T_{\bar{a}}A = AT_{\bar{a}}$ if and only if $A$ is an modulated circulant matrix.*

## 3. Modulated Quantum Fourier Transform

In this subsection, based on definition 2, we introduce one of the quantum primitives required for our algorithms.

**Definition 3.** *The modulated quantum Fourier transform (MQFT) associated with parameters $\bar{a} = (a_0, a_1, \ldots, a_{N-1})$ and an $N$-th root of $\prod_{j=0}^{N-1} a_j$ is the unitary operator*

$$\widetilde{F}_N : |x\rangle \;\longrightarrow\; \frac{1}{\sqrt{N}} \sum_{y=0}^{N-1} \gamma_y \, e^{2\pi i\, xy/N} |y\rangle, \quad (0 \le x \le N-1). \tag{11}$$

The standard quantum circuit for $\widetilde{F}_N$ is shown in Figure 1. It consists of Hadamard gates, controlled phase rotations $R_k = \begin{pmatrix} 1 & 0 \\ 0 & e^{2\pi i/2^k} \end{pmatrix}$, and final SWAP gates to reverse qubit order.

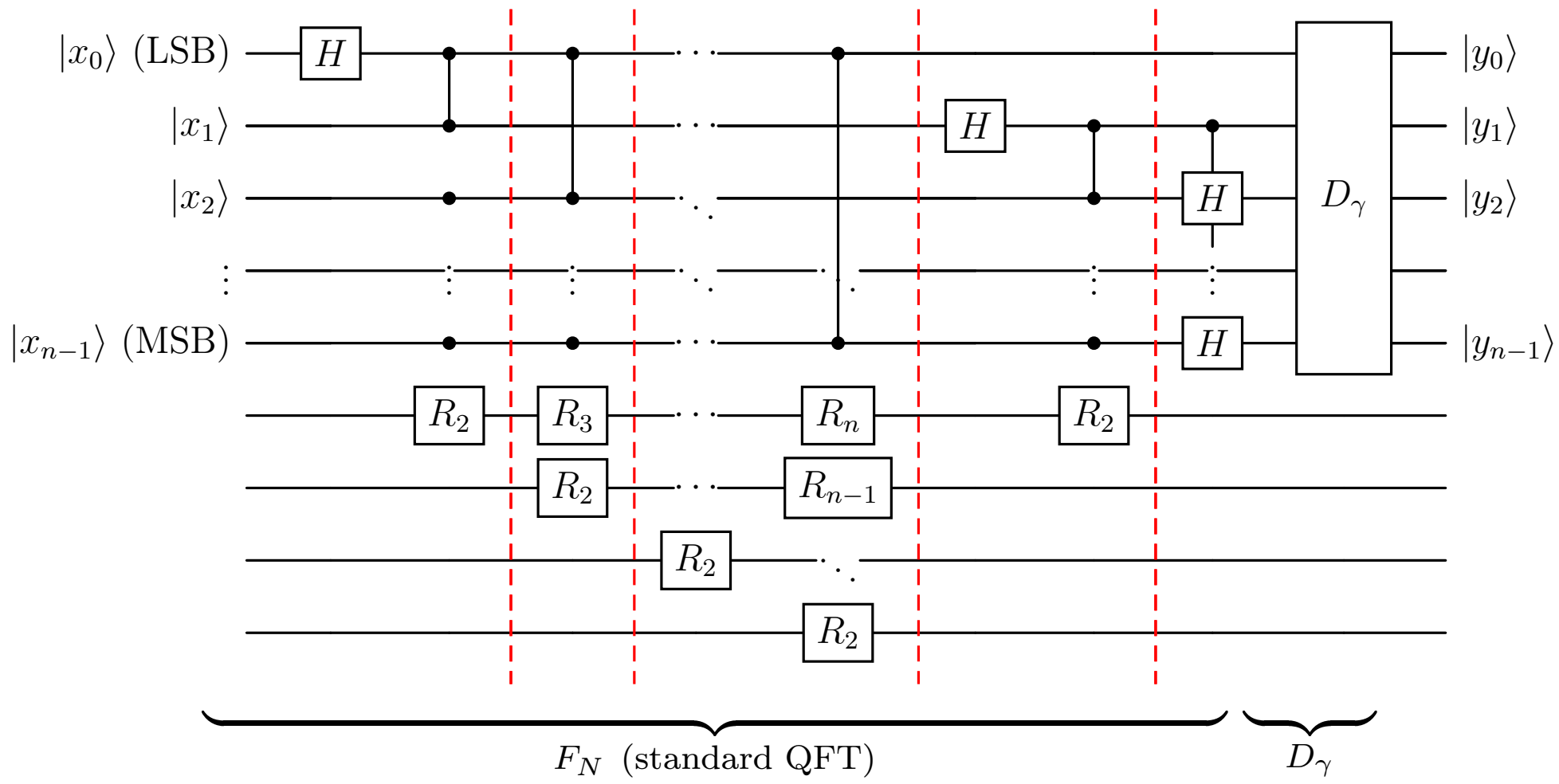

Figure 1: Quantum circuit for the Modulated Quantum Fourier Transform (MQFT) $\widetilde{F}_N = D_\gamma \cdot F_N$ on $n$ qubits ($N = 2^n$). The left block implements the standard QFT $F_N$: each qubit $|x_k\rangle$ first receives a Hadamard gate $H$, then controlled phase rotations $R_j = \left(\begin{smallmatrix} 1 & 0 \\ 0 & e^{2\pi i/2^j} \end{smallmatrix}\right)$ from the subsequent qubits. Vertical slices (dashed lines) mark the $n$ stages of the QFT. The right block applies the diagonal modulation gate $D_\gamma = \mathrm{diag}(\gamma_0, \ldots, \gamma_{N-1})$ with $\gamma_y = \gamma_{y-1} \frac{\gamma}{a_{y-1}}$, encoding the parameters $\bar{a} = (a_0, \ldots, a_{N-1})$ of the modulated circulant matrix (Definition 1). Under the oracle model of Remark 3, $D_\gamma$ is implementable in $O(\mathrm{poly}(n))$ gates, giving total MQFT complexity $O(n^2) = O(\log^2 N)$. The inverse MQFT reads $\widetilde{F}_N^\dagger = F_N^\dagger \cdot D_\gamma^\dagger$ (modulation applied *before* the inverse QFT). Output SWAP gates are omitted for clarity.

Thus, the total gate complexity of $\widetilde{F}_N$ (and its inverse) is $O(n^2) = O(\log^2 N)$, assuming efficient oracle access to the phase factors $\gamma_y$.

As in existing quantum algorithms for structured matrix–vector multiplication [7, 3], our method exploits the spectral structure of modulated circulant-type operators through the modulated quantum Fourier transform. To amplify the success probability of post-measurement outcomes, we employ amplitude amplification in the sense of Brassard et al. [8] , and efficient quantum arithmetic in [9]. Moreover, following standard assumptions in quantum linear algebra [1], we assume the efficient preparation of normalized quantum states encoding both the input vector and the first row of the structured matrix.

**Theorem 4.** *The modulated quantum Fourier transform $\widetilde{F}_N$ can be implemented with gate complexity $O(n^2)$, i.e., the same asymptotic complexity as the standard QFT.*

*Proof.* The standard QFT circuit produces the state $\frac{1}{\sqrt{N}} \sum_{y=0}^{N-1} e^{2\pi i\, xy/N}|y\rangle$. To obtain the modulation factors $\gamma_y$, we apply single-qubit phase gates $P(\phi_y)$ on each basis state $|y\rangle$, where $e^{i\phi_y} = \gamma_y$. All $\gamma_y$ are unimodular, and the phases can be applied efficiently in the Fourier basis using $O(n)$ controlled-phase gates or direct diagonal unitaries. Alternatively, the modulation can be implemented by applying a diagonal unitary with diagonal entries $\gamma_0, \gamma_1, \dots, \gamma_{N-1}$, after (or before) the standard QFT, which costs $O(n)$ gates in the worst case. Combined with the $O(n^2)$ cost of the standard QFT, the total complexity remains $O(n^2)$. □

**Remark 3.** *The $O(n^2)$ gate complexity stated in Theorem 4 relies on the structured form of the modulation phases $\gamma_y$, which satisfy the recursion $\gamma_y = \gamma_{y-1}\dfrac{\gamma}{a_{y-1}}; \gamma_0 = 1$ and can therefore be computed in $O(\text{poly}(n))$ gates. This stands in contrast to a* general *diagonal unitary with $N$ independent phases, which requires $\Omega(N)$ resources in the worst case [10]. In the numerical simulations of Section 4, $D_\gamma$ is synthesized by explicitly iterating over all $N$ computational basis states–an approach that is correct for small–scale validation but does not exploit this recursive structure; see the accompanying code for the explicit disclaimer.*

$$|x\rangle \text{ ($n$ qubits)} \;—\; \boxed{F_N} \;—\; \boxed{D_\gamma} \;—\; \widetilde{F}_N|x\rangle$$

Figure 2: Schematic circuit for the MQFT: $\widetilde{F}_N = D_\gamma \cdot F_N$. The standard QFT $F_N$ is applied first, followed by the diagonal phase gate $D_\gamma = \text{diag}(\gamma_0, \gamma_1, \dots, \gamma_{N-1})$ with entries $e^{i\phi_y}$ where $\phi_y = \arg(\gamma_y)$. The inverse MQFT is $\widetilde{F}_N^\dagger = F_N^\dagger \cdot D_\gamma^\dagger$ (phases applied before the inverse QFT).

The MQFT is the key tool that will allow us to extend known quantum algorithms for standard circulant matrices to the more general modulated circulant matrices in the subsequent sections.

**Example 3.** *Consider* $N = 3$*,* $\bar{a} = (a_0, a_1, a_2)$ *with* $a_i \in \mathbb{C} \setminus \{0\}$ *for all* $i$*, and* $\boldsymbol{x} = \mathbf{1}_3$*. Then, from Definition 1, the corresponding modulated circulant matrix is*

$$M_{\bar{a}}(\mathbf{1}) = \begin{bmatrix} 1 & a_0 & a_0a_1 \\ a_1a_2 & 1 & a_1 \\ a_2 & a_0a_2 & 1 \end{bmatrix}.$$

*This matrix is a* 3*-parametric circulant matrix and if* $|a_0| = |a_1| = |a_2| = 1$*,* $M_{\bar{a}}(\boldsymbol{1})$ *is a modulated circulant matrix.*

*To illustrate the difference with standard circulant matrices, we performed numerical simulations using Quantum Algorithm 2 from Hou et al. [7], which relies on the standard QFT and therefore diagonalizes only standard circulant matrices. The simulations were carried out with parameters* $a_0 = 0.5 + 0.3i$*,* $a_1 = 0.8 - 0.2i$*, and* $a_2 = 0.6 + 0.1i$*, normalized to unit magnitude:* $|\bar{a}| \approx [0.8575 + 0.5145i, 0.9701 - 0.2425i, 0.9864 + 0.1644i]$*) and input vector* $\boldsymbol{x} = (1, 0.5, 0.3)^T$*. Since quantum circuit implementations typically operate on power-of-two dimensions, the system was padded to* $N = 4$ *with identity parameters in the modulation vector.*

*Our MQFT-based algorithm yields a post-selection probability of approximately 0.483. The raw fidelity between the quantum output and the exact padded normalized result is 0.818. After optimal circular alignment (shift of 2 positions), the fidelity improves significantly to 0.990, with an L2 error reduced to 0.106, demonstrating that the modulated structure is correctly captured up to a global cyclic shift and minor phase offset.*

*In contrast, standard circulant multiplication via FFT (simulating the behavior of Algorithm 2 in [7], which relies on the standard QFT) achieves a fidelity of 0.991 with the quantum output. This close agreement in the padded case is attributed to the specific choice of constant coefficients* $\boldsymbol{x} = (1, 1, 1)$ *and the padding strategy with unit parameters, which partially restores standard circulant-like behavior. For more general parameter sets, non-constant* $\boldsymbol{x}$*, or non-padded simulations, larger discrepancies are expected between the standard QFT-based approach and the exact modulated result, highlighting the necessity and advantage of the MQFT for truly* $N$*-parametric modulated circulant matrices.*

*Figures 3–5 summarize the numerical results across all four test cases (*$N \in \{3, 4, 8\}$*, with zero-padding applied to* $N = 3$ *to reach the nearest power of two). Figure 3 shows the fidelity of the quantum output state with respect to both the exact modulated result* $M_{\bar{a}}(\boldsymbol{x})\,v$ *and the standard circulant product computed via FFT. For the two cases without padding (*$N = 4$ *and*

*$N = 8$), the MQFT-based algorithm achieves fidelities of $0.973$ and $0.981$, respectively, confirming the correctness of the proposed circuit. The lower fidelities observed for $N = 3$ are attributable solely to the zero-padding strategy ($N = 3 \to 4$), which modifies the effective matrix structure; this artifact vanishes when $N$ is already a power of two.*

*Figure 4 compares the theoretical success probability*

$$P_{\text{succ}} = \frac{1}{\kappa^2} \sum_{k=0}^{N-1} |\beta_k|^2 |\mu_k|^2, \qquad \kappa = \max_k |\mu_k|, \tag{12}$$

*with the value measured in the Qiskit simulation. For the non-padded cases, the agreement is excellent: relative errors of $3.5\%$ ($N = 4$) and $0.16\%$ ($N = 8$), validating formula (12). For the padded cases ($N = 3$, shown with hatching in Figure 4), a large discrepancy is observed and isexpected: formula (12) is evaluated using the eigenstructure of the* original *$3 \times 3$ matrix, whereas the quantum simulation necessarily operates on the padded $4$-dimensional system. These correspond to systems of different effective dimension—the padding introduces additional basis states and modifies the eigenvalue spectrum—so a direct numerical comparison of $P_{\text{succ}}$ across dimensions is not meaningful. The fidelity values (Figure 3), which compare quantum output states* within the same *($N = 4$) Hilbert space, remain well-defined and confirm that the circuit correctly implements the padded modulated circulant multiplication. Finally, Figure 5 displays the eigenvalue magnitudes $|\mu_j|$ of $M_{\bar{a}}(\boldsymbol{x})$ for each test case. The pronounced non-uniformity of the spectrum—with $|\mu_0|$ significantly larger than the remaining eigenvalues—illustrates why the standard QFT is insufficient here: the modulation factors $\gamma_y$ are non-trivial and case-dependent, making the MQFT essential for exact diagonalization.*

## 4. Quantum circuits

In this section, we present efficient quantum circuits for implementing key operations associated with modulated circulant matrices, explicitly exploiting their spectral decomposition via modulated quantum Fourier transform introduced earlier. A central building block of any modulated circulant matrix $M_{\bar{a}}(x)$ is the modulated cyclic permutation matrix $T_{\bar{a}}$, defined in Equation (3). This matrix plays a role analogous to the standard cyclic shift operator in conventional circulant matrices: every modulated circulant matrix can be expressed as a polynomial in $T_{\bar{a}}$ with coefficients given by the vector $x$.

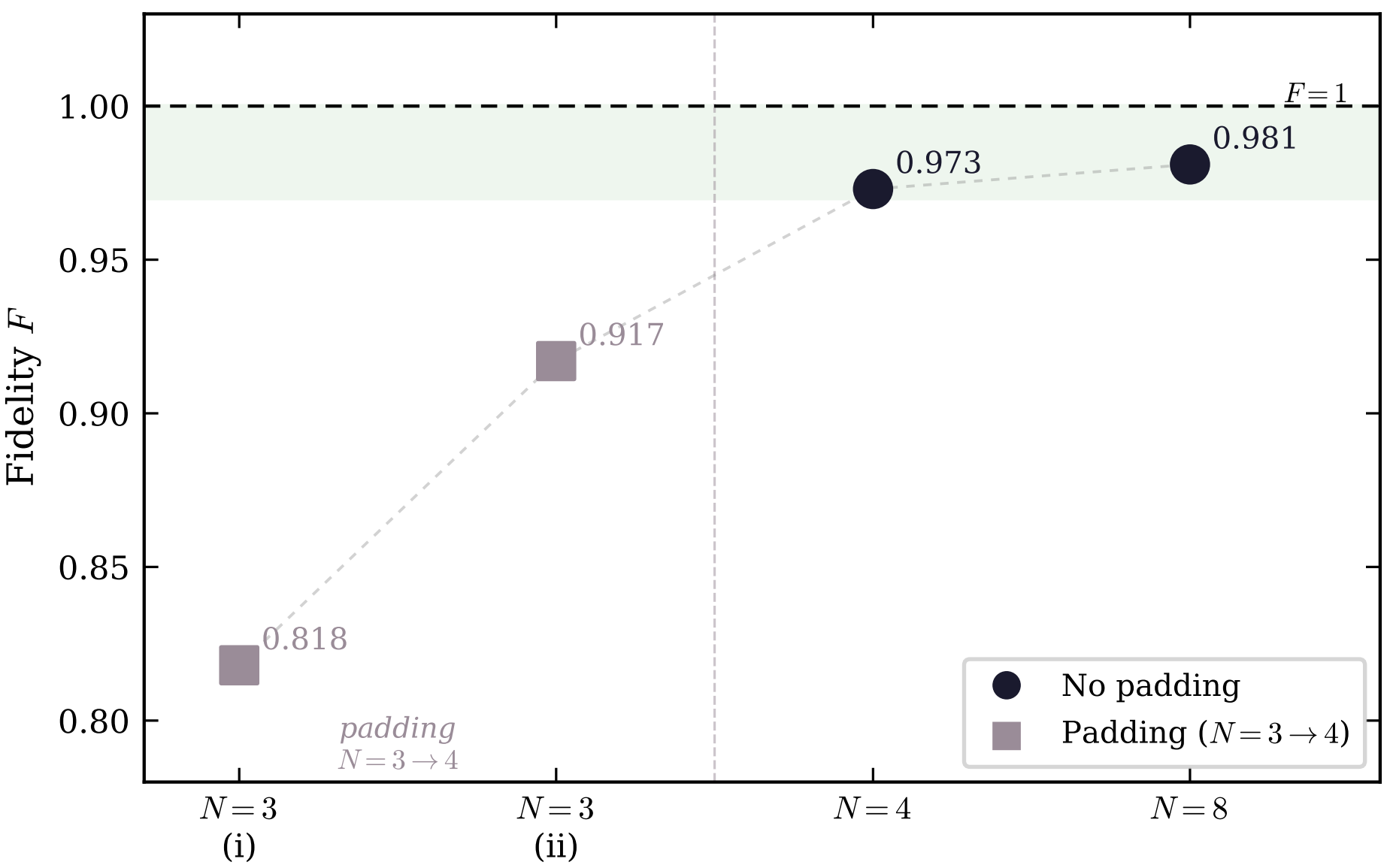

Figure 3: Fidelity $|\langle \psi_{\text{out}} | \psi_{\text{exact}} \rangle|^2$ of the quantum output state with respect to the exact modulated circulant product $M_{\bar{a}}(\tilde{c})\, v$, for each test case. Dashed lines indicate reference thresholds at $F = 0.99$ and $F = 0.90$. Hatched bars correspond to cases requiring zero-padding ($N = 3 \to 4$); in these cases the circuit operates in the padded dimension $N = 4$ while the target product is defined for $N = 3$.

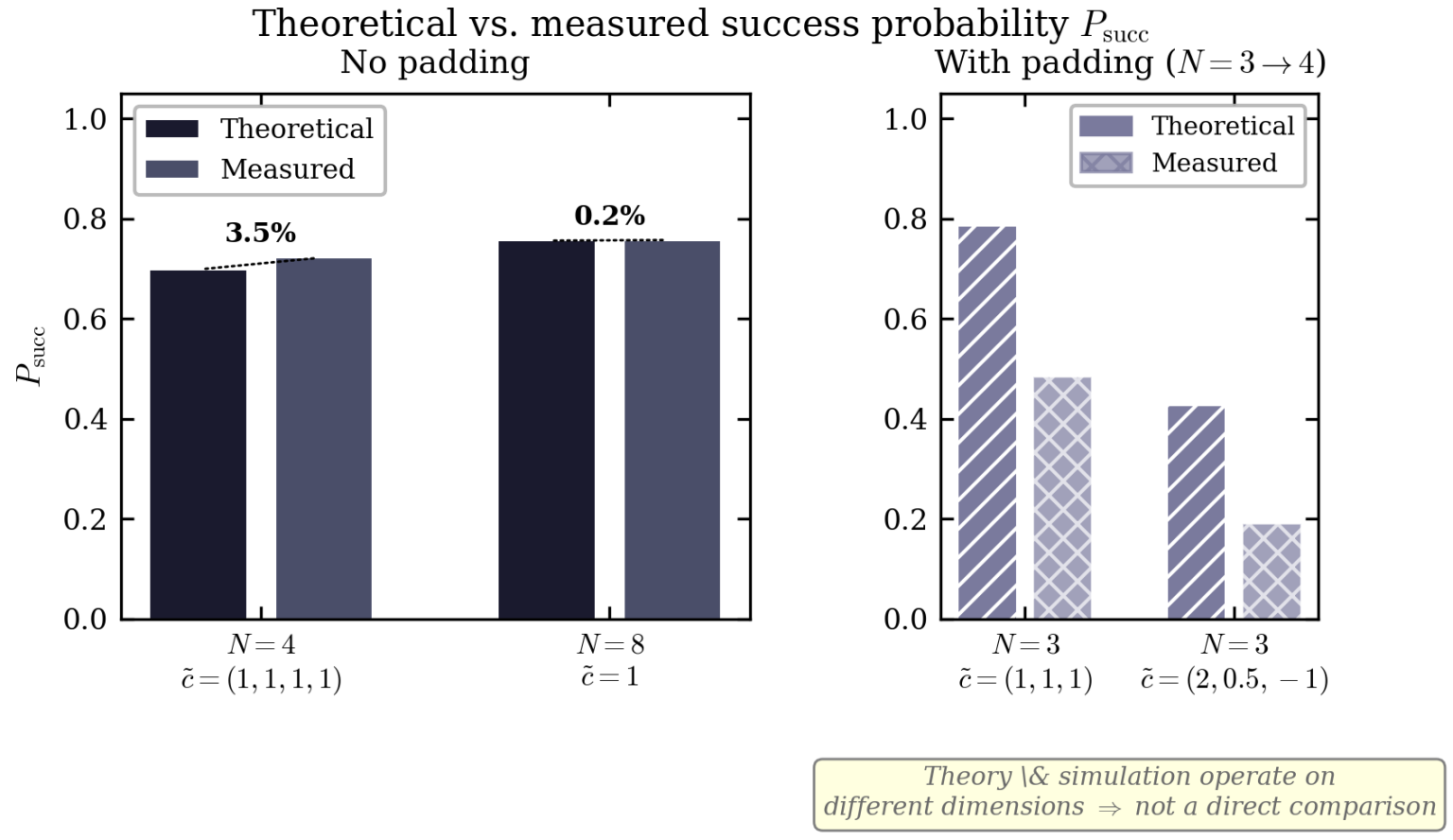

Figure 4: Theoretical success probability $P_{\text{succ}}$ (Eq. (12), dark bars) versus the value measured in the Qiskit simulation (gray bars), for each test case. Relative errors are annotated above the bar pairs for the non-padded cases ($N = 4$: 3.5%; $N = 8$: 0.16%). Hatched bars indicate padded cases ($N = 3 \to 4$) where the theoretical formula and the simulation operate on systems of different effective dimension; the observed discrepancy is therefore expected and does not reflect an error in the algorithm (see text for a detailed discussion).

Efficient quantum implementation of $T_{\bar{a}}$ is therefore crucial, as repeated controlled applications of this operator will allow us to evaluate the complete matrix vector product $M_{\bar{a}}(x)|v\rangle$ with polylogarithmic complexity in the matrix order $N = 2^n \times 2^n$, assuming efficient state preparation and oracle access to the parameters. Moreover, since $T_{\bar{a}}$ is unitary it admits an efficient and exact unitary quantum circuit, without the need for approximate techniques such as block-encoding.

### *4.1. Quantum Circuit for the Controlled-$T_{\bar{a}}$ Transformation*

We present an efficient quantum circuit for implementing the modulated cyclic permutation matrix $T_{\bar{a}}$ on $n$ qubits ($N = 2^n$). From the spectral decomposition (7),

$$T_{\bar{a}} = F_{\bar{a}}\, \Lambda_\gamma\, F_{\bar{a}}^{\dagger}, \tag{13}$$

where $\Lambda_\gamma = \operatorname{diag}(\gamma, \gamma\omega, \ldots, \gamma\omega^{N-1})$.

Since the matrix representation of the MQFT $\widetilde{F}_N = D_\gamma F_N$ in the computational basis coincides with $F_{\bar{a}}$ (Definition 3), the spectral decomposi-

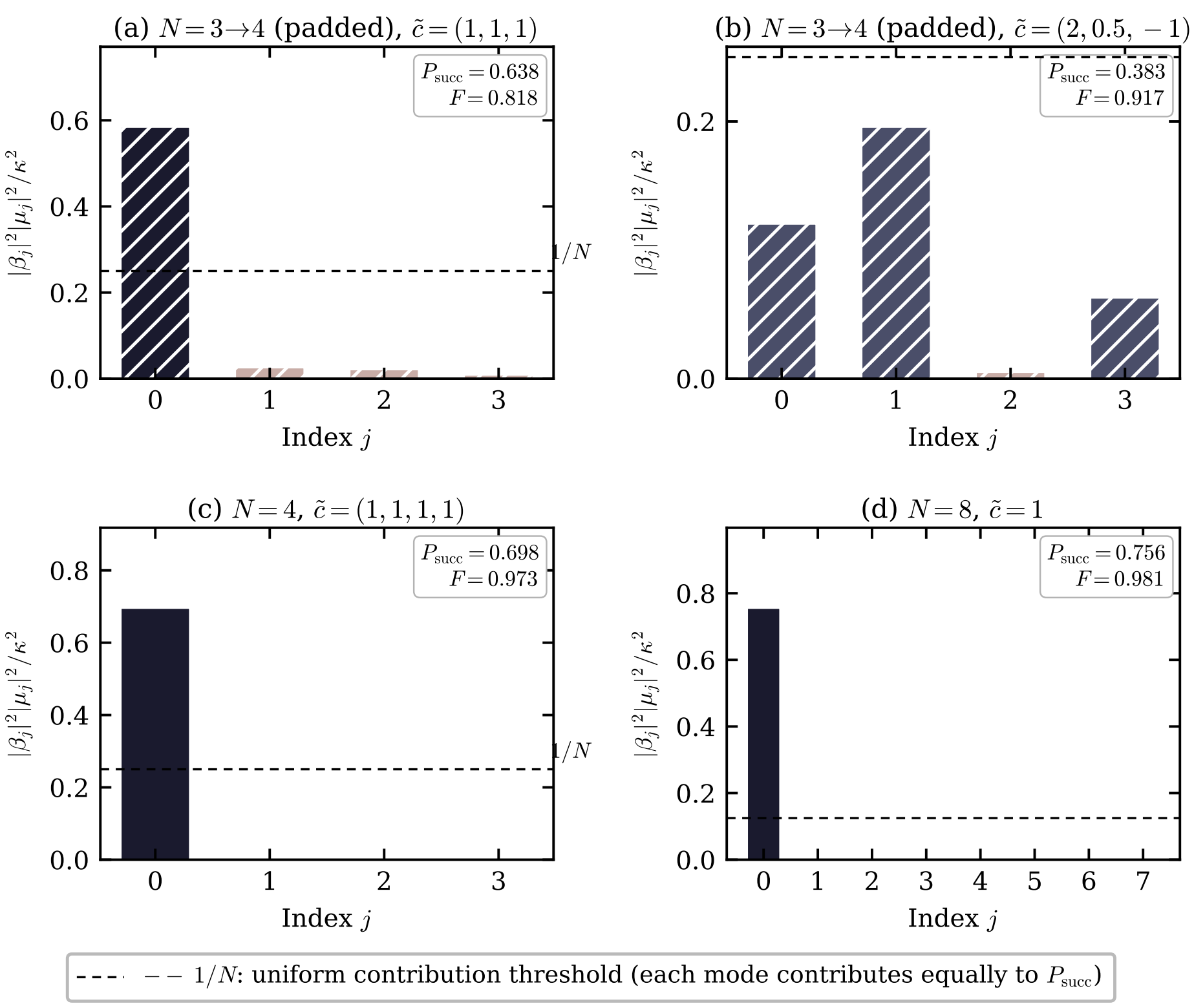

Figure 5: Per-mode contributions $|\beta_j|^2|\mu_j|^2\kappa^2$ to the success probability $P_{\rm succ}$, for each test case. Panels (a) and (b) correspond to $N = 3$ padded to $N = 4$ for the quantum simulation; the four bars therefore represent the padded space, not the original $N = 3$ problem. The dashed line marks the uniform threshold $1/N$: if all modes contributed equally, each would carry exactly $1/N$; modes exceeding this threshold dominate the sum $P_{\rm succ} = \sum_j |\beta_j|^2|\mu_j|^2/\kappa^2$. In cases without padding (panels c–d) a single mode concentrates most of the weight, explaining the higher $P_{\rm succ}$. Hatched bars indicate padded dimensions.

$$|x\rangle \;—\; \boxed{\widetilde{F}_N^\dagger} \;—\; \boxed{\Lambda_\gamma} \;—\; \boxed{\widetilde{F}_N} \;—\; T_{\bar{a}}|x\rangle$$

Figure 6: Quantum circuit implementing $T_{\bar{a}} = \widetilde{F}_N \Lambda_\gamma \widetilde{F}_N^\dagger$. The inverse MQFT $\widetilde{F}_N^\dagger$ is applied first to map the input to the modulated Fourier basis; the diagonal gate $\Lambda_\gamma$ applies the phase $\gamma\omega^k$ to each basis state $|k\rangle$; finally $\widetilde{F}_N$ returns to the computational basis. In the simulation code, $\Lambda_\gamma$ is synthesized by explicitly enumerating all $N$ basis states (cost $O(N \cdot \text{poly}(n))$), but exploiting the recursive structure of $\gamma_y$ (Remark 3) reduces this to $O(\text{poly}(n))$ gates. The total circuit cost is therefore $O(n^2) = O(\log^2 N)$.

tion (13) can be written equivalently in terms of the MQFT as

$$T_{\bar{a}} = \widetilde{F}_N \, \Lambda_\gamma \, \widetilde{F}_N^\dagger. \tag{14}$$

This decomposition leads to the following three-step implementation:

1. Apply $\widetilde{F}_N^\dagger$ to express the input state in the modulated Fourier basis, i.e., the eigenbasis of $T_{\bar{a}}$.
2. Apply the diagonal unitary $\Lambda_\gamma = \text{diag}(\gamma\omega^0, \gamma\omega^1, \ldots, \gamma\omega^{N-1})$, which acts as
$$\Lambda_\gamma \, |k\rangle \;\mapsto\; \gamma\omega^k \, |k\rangle, \qquad k = 0, \ldots, N-1.$$
   This multiplies each amplitude by the corresponding eigenvalue of $T_{\bar{a}}$ in the modulated Fourier basis.
3. Apply $\widetilde{F}_N$ to return to the computational basis.

The resulting circuit is shown in Figure 6.

The gate complexity is dominated by the two MQFT applications. From Theorem 4, each $\widetilde{F}_N$ and its inverse costs $O(n^2) = O(\log^2 N)$ gates. The diagonal gate $\Lambda_\gamma$ can be implemented efficiently in two ways:

- *Via diagonal-unitary synthesis.* The phases $\gamma\omega^k$ are classically computable from $\bar{a}$ and $\gamma$; standard techniques [10, 11] implement $\Lambda_\gamma$ with $O(\text{poly}(n))$ gates and bounded error $\epsilon$.

- *Via single-qubit phase gates.* Since $\omega^k = e^{2\pi i k/N}$ and $k = \sum_{b=0}^{n-1} k_b 2^b$ in binary, the phase $\gamma\omega^k$ factorizes as $\gamma \prod_{b=0}^{n-1} e^{2\pi i k_b 2^b/N}$. Hence, up to the irrelevant global phase $\gamma$, $\Lambda_\gamma$ can be implemented using $n$ single-qubit phase gates, one per qubit, contributing only $O(n)$ additional gates.

In either case $\Lambda_\gamma$ costs at most $O(\text{poly}(n))$ gates, and the total complexity for implementing $T_{\bar{a}}$ is $O(n^2) = O(\log^2 N)$.

**Remark 4.** *According to Definition 1, $|a_j| = 1$ for all $j$, then $\gamma = \prod_{j=0}^{N-1} a_j$ has unit modulus. Thus, $\Lambda_\gamma$ is unitary and the circuit of Figure 6 is an exact unitary quantum circuit, requiring no block-encoding or non-unitary approximation techniques.*

The implementation of $T_{\bar{a}}$ serves as the foundational subroutine for the matrix–vector multiplication algorithm presented in section 5.

### *4.2. Quantum Circuit for the Controlled-$T_{\bar{a}}$ Transformation*

In this subsection we introduce the quantum circuit for the controlled application of powers of the modulated cyclic permutation $T_{\bar{a}}$, which appears as a key sub-circuit in quantum phase estimation procedures applied to modulated circulant matrices. Given an state $|a\rangle$ we wish to evolve under powers of $T_{\bar{a}}$, the transformation implemented by the circuit is

$$|0\rangle^{\otimes n}|a\rangle \;\longrightarrow\; \frac{1}{\sqrt{2^n}} \sum_{i=0}^{2^n-1} |i\rangle\, T_{\bar{a}}^i\, |a\rangle. \tag{15}$$

This circuit, shown in Figure 7, consists of $n$ Hadamard gates on the ancilla (counting) register, followed by $n$ controlled-$T_{\bar{a}}^{2^k}$ operations acting on the target register.

This circuit requires $n$ controlled operations of the form controlled-$T_{\bar{a}}^{2^k}$. The implementation cost of each controlled-$T_{\bar{a}}^{2^k}$ depends on the circuit realization of $T_{\bar{a}}$ presented in the previous subsection. Since $T_{\bar{a}}$ can be implemented with $O(n^2)$ gates, and exponentiation by $2^k$ can be achieved via repeated squaring or direct phase adjustment in the diagonal basis, the total gate complexity remains polynomial in $n$.

**Proposition 1.** *The circuit of Figure 7 can be implemented with gate complexity $O(n^3) = O(\log^3 N)$.*

*Proof.* The circuit implementation requires $n$ controlled-$T_{\bar{a}}^{2^k}$ operations, for $k = 0, 1, \ldots, n-1$. Each power $T_{\bar{a}}^{2^k}$ is computed in the diagonal basis; according to Equation (14), where $T_{\bar{a}} = \mathcal{F}_N^\dagger \Lambda_\gamma \mathcal{F}_N$, it follows that:

$$T_{\bar{a}}^{2^k} = \widetilde{F}_N^\dagger \Lambda_\gamma^{2^k} \widetilde{F}_N, \tag{16}$$

where $\Lambda_\gamma^{2^k} = \mathrm{diag}(\gamma^{2^k},\, \gamma^{2^k}\omega^{2^k},\, \ldots,\, \gamma^{2^k}\omega^{(N-1)2^k})$. The operator $\Lambda_\gamma^{2^k}$ preserves the recursive structure of $\Lambda_\gamma$ and can be implemented using $O(\mathrm{poly}(n))$ gates (see Remark 3). Consequently, each controlled-$T_{\bar{a}}^{2^k}$ operation incurs a cost of $O(n^2)$ gates, primarily dominated by the two MQFT applications. Given $n$ such operations, the total computational complexity is $O(n \cdot n^2) = O(n^3) = O(\log^3 N)$. $\square$

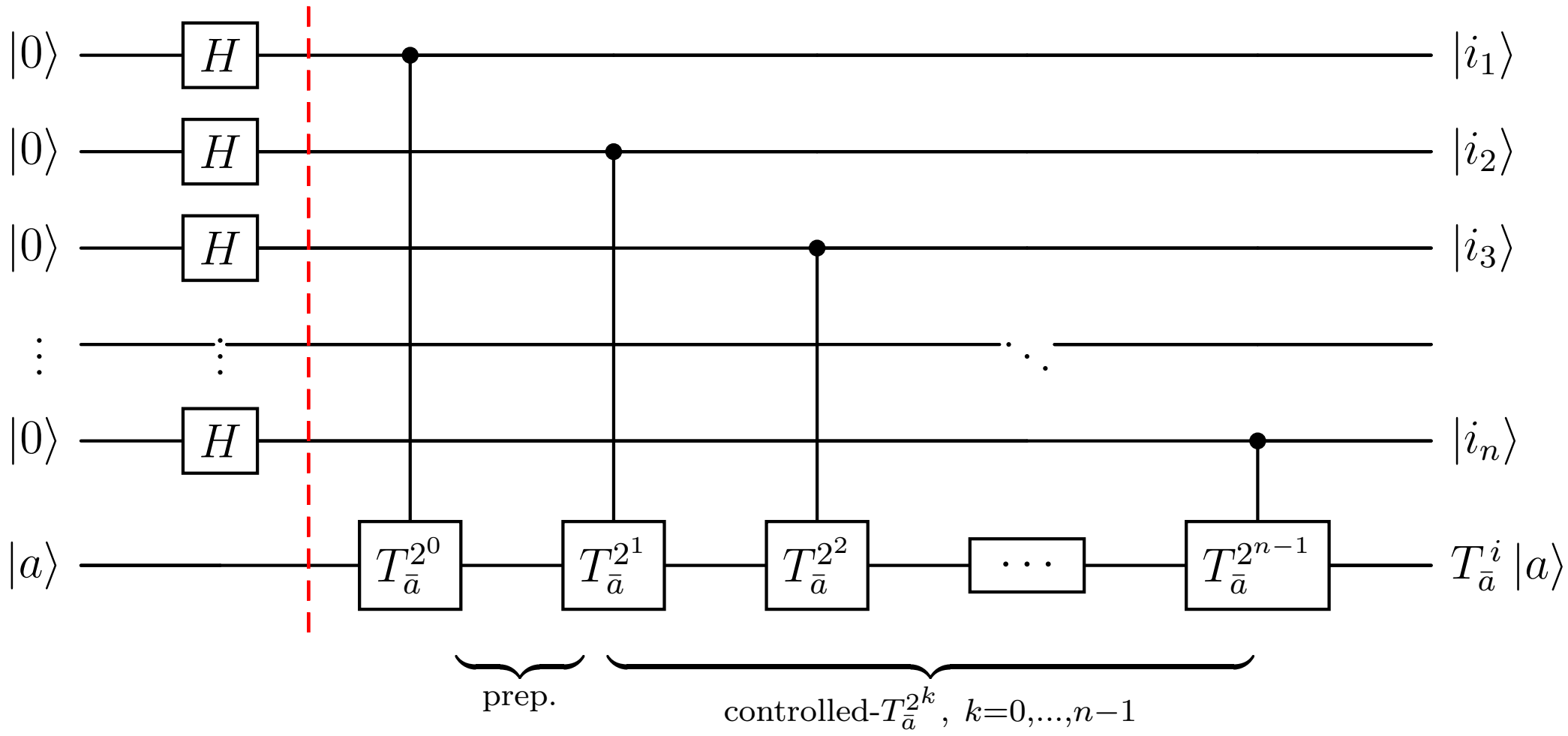

Figure 7: Quantum circuit implementing the uniform superposition over all powers of $T_{\bar{a}}$. The top $n$ qubits (ancilla/counting register) are each prepared in $|+\rangle = H|0\rangle$; vertical slices mark the Hadamard stage. The bottom qubit (target register), initially $|a\rangle$, receives $n$ successive controlled-$T_{\bar{a}}^{2^k}$ operations, one per ancilla qubit. The output state is $\frac{1}{\sqrt{N}}\sum_{i=0}^{N-1}|i\rangle\, T_{\bar{a}}^i\,|a\rangle$ (with $N = 2^n$), which encodes all polynomial evaluations needed for $M_{\bar{a}}(\mathbf{x})\,|a\rangle$ via equation (9). Each controlled-$T_{\bar{a}}^{2^k}$ is implemented using the spectral decomposition $T_{\bar{a}}^{2^k} = \widetilde{F}_N^\dagger\,\Lambda_\gamma^{2^k}\,\widetilde{F}_N$ at cost $O(n^2)$ gates (Proposition 1).

## 5. Multiplication of Modulated Circulant Matrices and Vectors

In this section, we introduce a quantum algorithm for computing the product of a modulated circulant matrix $M_{\bar{a}}(x)$ and an input vector $v$, both encoded as quantum states. The algorithm leverages the spectral decomposition of $M_{\bar{a}}(x)$ via the MQFT. It yields the normalized result as a quantum state, conditioned upon a post-selection event whose probability is analytically computable.

### 5.1. Three-register circuit and algorithm

The algorithm uses the following three $n$-qubit registers:

- $\mathcal{E}$: the *encoding register*, initialized with the normalized eigenvalue vector of $M_{\bar{a}}(\mathbf{x})$;
- $\mathcal{X}$: the *input register*, initialized with the normalized input vector $v$;
- $\mathcal{A}$: the *ancilla register*, used to implement the post-selection via a bitwise comparison.

**Input preparation.** From Corollary 1, the eigenvalues of $M_{\bar{a}}(\mathbf{x})$ are $\mu_j = (F_{\bar{a}}^{\dagger}\mathbf{x})_j$, $j = 0, \ldots, N-1$. Set $\kappa = \max_j |\mu_j|$ and define the normalized eigenvalue state

$$|\mu\rangle_{\mathcal{E}} = \frac{1}{\kappa\sqrt{N}} \sum_{k=0}^{N-1} \mu_k \, |k\rangle, \tag{17}$$

and the normalized input state $|v\rangle_{\mathcal{X}} = \|v\|^{-1} \sum_{k=0}^{N-1} v_k \, |k\rangle$. The initial $3n$-qubit state is

$$|\Psi_0\rangle = |\mu\rangle_{\mathcal{E}} \otimes |v\rangle_{\mathcal{X}} \otimes |0\rangle_{\mathcal{A}}^{\otimes n}. \tag{18}$$

**Step 1: Apply $\widetilde{F}_N^{\dagger}$ to $\mathcal{X}$.** The inverse MQFT is applied to the input register, thereby expressing $|v\rangle_{\mathcal{X}}$ in the modulated Fourier basis. Since the matrix representation of $\widetilde{F}_N^\dagger$ in the computational basis coincides with $F_{\bar{a}}^\dagger$ (Definition 3), the resulting amplitudes are precisely $\beta_k = (F_{\bar{a}}^\dagger v/\|v\|)_k$:

$$|\Psi_1\rangle = \frac{1}{\kappa N\|v\|} \sum_{j=0}^{N-1} \sum_{k=0}^{N-1} \mu_j \, \beta_k \, |j\rangle_{\mathcal{E}} \otimes |k\rangle_{\mathcal{X}} \otimes |0\rangle_{\mathcal{A}}^{\otimes n}. \tag{19}$$

**Step 2: XOR into the ancilla.** Two layers of CNOT gates compute the bitwise XOR of the $\mathcal{X}$-index and the $\mathcal{E}$-index into $\mathcal{A}$: for each bit $b = 0, \ldots, n-1$, first $\text{CNOT}(\mathcal{X}_b \to \mathcal{A}_b)$, then $\text{CNOT}(\mathcal{E}_b \to \mathcal{A}_b)$. After this step,

$$|\Psi_2\rangle = \frac{1}{\kappa N\|v\|} \sum_{j,k} \mu_j \, \beta_k \, |j\rangle_{\mathcal{E}} \otimes |k\rangle_{\mathcal{X}} \otimes |j \oplus k\rangle_{\mathcal{A}}. \tag{20}$$

**Step 3: Post-selection on $\mathcal{A} = |0\rangle^{\otimes n}$.** Retaining only the terms where $j \oplus k = 0$, i.e., $j = k$, the post-selected (unnormalized) state is

$$|\Psi_3\rangle \propto \sum_{k=0}^{N-1} \mu_k \, \beta_k \, |k\rangle_{\mathcal{E}} \otimes |k\rangle_{\mathcal{X}}, \tag{21}$$

and the probability of this outcome is

$$P_{\text{succ}} = \frac{1}{\kappa^2} \sum_{k=0}^{N-1} |\beta_k|^2 \, |\mu_k|^2, \tag{22}$$

which coincides with formula (12).

**Step 4: Uncompute $\mathcal{X}$.** A layer of CNOT gates from $\mathcal{E}$ to $\mathcal{X}$ disentangles the two registers:

$$|\Psi_4\rangle \propto \left( \sum_{k=0}^{N-1} \mu_k \, \beta_k \, |k\rangle \right)_{\mathcal{E}} \otimes \; |0\rangle_{\mathcal{X}}^{\otimes n}. \tag{23}$$

**Step 5: Apply $\widetilde{F}_N$ to $\mathcal{E}$.** The MQFT on $\mathcal{E}$ maps the diagonal product back to the computational basis. From Corollary 1, $M_{\bar{a}}(\mathbf{x}) = F_{\bar{a}} \, \text{diag}(\mu_0, \ldots, \mu_{N-1}) \, F_{\bar{a}}^\dagger$, so identifying $\widetilde{F}_N = F_{\bar{a}}$:

$$\widetilde{F}_N \sum_{k=0}^{N-1} \mu_k \, \beta_k \, |k\rangle \;=\; \widetilde{F}_N \, \text{diag}(\mu_0, \ldots, \mu_{N-1}) \, \widetilde{F}_N^\dagger \, |v\rangle \;\propto\; M_{\bar{a}}(\mathbf{x}) \, v, \tag{24}$$

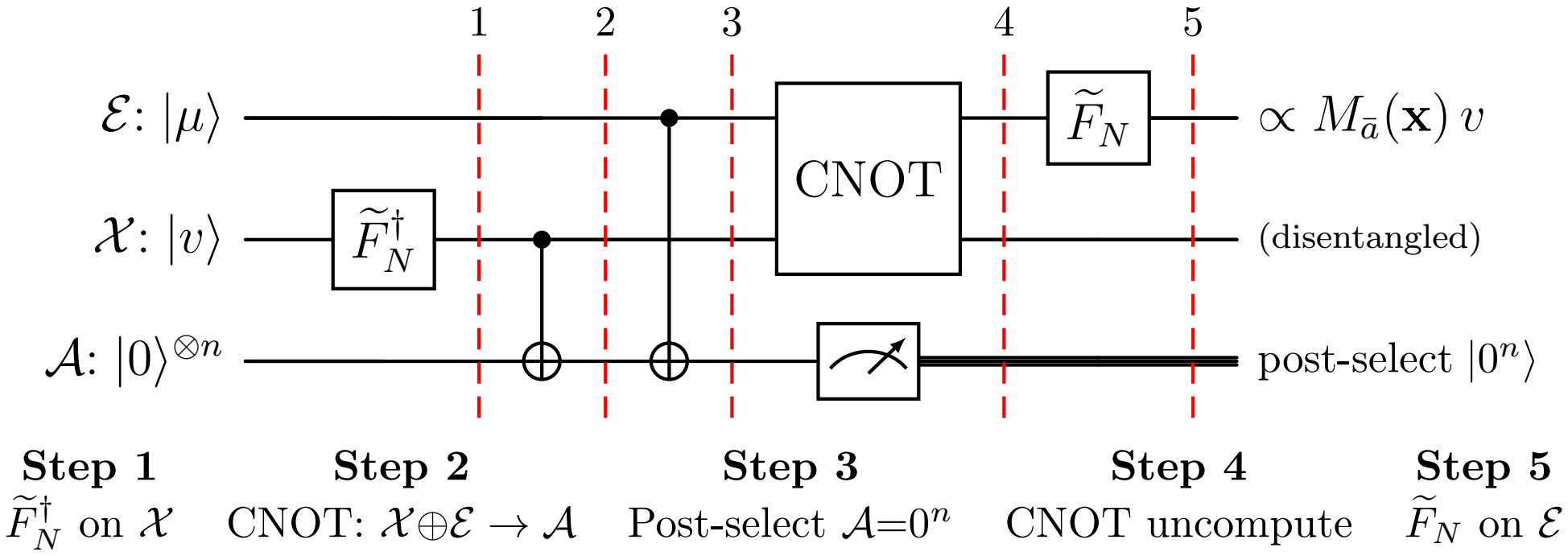

Figure 8: Quantum circuit for the MQFT-based matrix–vector multiplication algorithm (Section 5), using three $n$-qubit registers: $\mathcal{E}$ (eigenvalue encoding), $\mathcal{X}$ (input vector $v$), and $\mathcal{A}$ (ancilla). **Step 1**: apply $\widetilde{F}_N^\dagger$ to $\mathcal{X}$, expanding $|v\rangle$ in the modulated Fourier basis with coefficients $\beta_k = (F_{\bar{a}}^\dagger v/\|v\|)_k$. **Step 2**: two CNOT layers compute $|\mathcal{X}\rangle \oplus |\mathcal{E}\rangle$ into $\mathcal{A}$, thereby encoding the equality condition $j = k$ in the ancilla register. **Step 3**: measure $\mathcal{A}$ and post-select on $|0^n\rangle$, projecting onto the $j$=$k$ terms with success probability $P_{\text{succ}} = \kappa^{-2}\sum_k |\beta_k|^2|\mu_k|^2$ (Theorem 5). **Step 4**: CNOT gates from $\mathcal{E}$ to $\mathcal{X}$ disentangle the registers (uncompute). **Step 5**: apply $\widetilde{F}_N$ to $\mathcal{E}$; by Corollary 1, the output state is proportional to $M_{\bar{a}}(\mathbf{x})\, v \,/\, \|M_{\bar{a}}(\mathbf{x})\, v\|$.

so the final state of $\mathcal{E}$ encodes the normalized product $M_{\bar{a}}(\mathbf{x})\, v \,/\, \|M_{\bar{a}}(\mathbf{x})\, v\|$.

The complete circuit is shown in Figure 8.

The superposition state (15) explicitly encodes the information required to evaluate $M_{\bar{a}}(x)|a\rangle$ using the aforementioned algorithm. Given that the eigenvalues of $T_{\bar{a}}$ are $\gamma\omega^j$ for $j = 0, \ldots, N-1$, each term $T_{\bar{a}}^r|a\rangle$ contributes a phase factor $(\gamma\omega^j)^r$ within the modulated Fourier basis. By adding $r$ with weights $x_r$, one recovers the eigenvalues $\mu_j = \sum_{r=0}^{N-1} x_r(\gamma\omega^j)^r$ of $M_{\bar{a}}(x)$, consistent with the definition provided in Eq. (9).

### *5.2. Complexity and success probability*

**Theorem 5.** *The MQFT-based matrix–vector multiplication algorithm outputs the normalized product $M_{\bar{a}}(\boldsymbol{x})\, v/\|M_{\bar{a}}(\boldsymbol{x})\, v\|$ as a quantum state, with success probability*

$$P_{\text{succ}} = \frac{1}{\kappa^2} \sum_{k=0}^{N-1} |\beta_k|^2\, |\mu_k|^2,$$

*using $O(n^2) = O(\log^2 N)$ gates per trial, where $\kappa = \max_j |\mu_j|$.*

*Proof.* The computational complexity of the proposed algorithm is summarized as follows. Each of the two MQFT applications (Steps 1 and 5) requires $O(n^2)$ gates, by Theorem 4. The CNOT layers (Steps 2 and 4) involve $O(n)$ gates each. Under standard amplitude-encoding assumptions [1], the state preparation of $|\mu\rangle$ and $|v\rangle$ entails $O(N)$ classical pre-processing and $O(n)$ quantum gates. Consequently, the dominant cost is therefore $O(n^2) = O(\log^2 N)$. Finally, the expression for $P_{\text{succ}}$ is derived directly from the probability amplitude of the post-selected state at Step 3. □

**Remark 5.** *If $P_{\text{succ}}$ is small, amplitude amplification [8] boosts the success probability to $\Theta(1)$ using $O(1/\sqrt{P_{\text{succ}}})$ repetitions, preserving the $O(\log^2 N)$ per-query complexity whenever $P_{\text{succ}} = \Omega(1/\text{poly}(n))$.*

**Remark 6.** *When all entries of $\boldsymbol{x}$ and all amplitudes of $v$ are real and non-negative, the dominant eigenvalue satisfies $|\mu_0| = \kappa$ and the corresponding coefficient satisfies $|\beta_0|^2 \approx 1$, so $P_{\text{succ}} \approx 1$. In this regime no amplitude amplification is needed, and the algorithm achieves an exponential speedup over classical FFT-based methods, which require $O(N \log N)$ operations.*

### *5.3. Numerical validation*

To assess the correctness of the proposed MQFT-based matrix–vector multiplication algorithm, we implemented the circuit in Qiskit and evaluated it on four representative test cases. The numerical validation is intended to address two questions: first, whether the quantum circuit reproduces the normalized product $M_{\bar{a}}(\mathbf{x})v/\|M_{\bar{a}}(\mathbf{x})v\|$ with high fidelity; and second, whether the analytical expression for the post-selection success probability,

$$P_{\text{succ}} = \frac{1}{\kappa^2} \sum_{k=0}^{N-1} |\beta_k|^2 |\mu_k|^2, \qquad \kappa = \max_{0 \le k \le N-1} |\mu_k|, \tag{25}$$

accurately predicts the behavior observed in simulation.

Table 1 summarizes the theoretical success probability, the post-selection probability measured in Qiskit, their relative error, and the fidelity between the quantum output and the exact classical target state.

The clearest evidence is provided by the non-padded cases, namely $N = 4$ and $N = 8$. In these instances, the theoretical model and the quantum simulation are defined on the same Hilbert space, so both the fidelity and the post-selection probability can be compared directly. The measured fidelities, 0.973 and 0.981, indicate that the circuit accurately prepares the target output state. At the same time, the relative errors in $P_{\text{succ}}$ are only 3.5% and 0.2%, respectively, which supports the validity of (25) as a quantitative prediction of the algorithmic success probability.

Table 1: Numerical results for the MQFT-based algorithm across four test cases. $P_{\text{succ}}^{\text{theo}}$ is evaluated from (25) at the original dimension $N$, whereas $P_{\text{succ}}^{\text{meas}}$ is the post-selection probability measured in Qiskit. For the padded cases ($N = 3 \to 4$), the discrepancy in $P_{\text{succ}}$ is expected, since theory and simulation refer to systems of different effective dimension.

| Test case | $P_{\text{succ}}^{\text{theo}}$ | $P_{\text{succ}}^{\text{meas}}$ | Rel. error (%) | Fidelity |
|---|---|---|---|---|
| $N = 3$, $\mathbf{x} = (1, 1, 1)$ | 0.786 | 0.483 | 38.5 | 0.818 |
| $N = 3$, $\mathbf{x} = (2, 0.5, -1)$ | 0.426 | 0.190 | 55.4 | 0.917 |
| $N = 4$, $\mathbf{x} = (1, 1, 1, 1)$ | 0.698 | 0.722 | 3.5 | 0.973 |
| $N = 8$, $\mathbf{x} = \mathbf{1}$ | 0.756 | 0.757 | 0.2 | 0.981 |

The situation is different for the cases with original dimension $N = 3$, which must be embedded into dimension 4 in order to fit the qubit-based implementation. In these examples, the theoretical quantity $P_{\text{succ}}^{\text{theo}}$ is computed from the spectral data of the original $3 \times 3$ modulated circulant matrix, whereas the simulation is necessarily carried out on the padded 4-dimensional system. Therefore, the two values do not correspond to the same operator and should not be interpreted as directly comparable. From this viewpoint, the large discrepancies observed in $P_{\text{succ}}$ for the padded cases are an expected consequence of the embedding procedure rather than a deficiency of the MQFT-based construction.

For the padded cases, the fidelity remains the most meaningful validation metric, since it compares the simulated output state with the exact target state within the same implemented Hilbert space. Although the values are lower than in the non-padded regime, they still indicate that the circuit captures the intended modulated structure with reasonable accuracy after embedding. This distinction between padded and non-padded instances is also reflected in Figures 3 and 4: the former provides state-level validation through fidelity, whereas the latter confirms that the theoretical expression for $P_{\text{succ}}$ is reliable precisely when no dimensional mismatch is introduced.

Finally, Figure 5 helps interpret these results from a spectral viewpoint. The non-uniform distribution of the eigenvalue magnitudes $|\mu_j|$ explains the observed concentration of the success probability and illustrates the role of the modulation in the diagonalization process. In particular, these plots reinforce that the MQFT is not merely a reformulation of the standard QFT, but the appropriate transform for the exact treatment of modulated circulant matrices.

Overall, the numerical experiments should be understood as a proof-of-concept validation of the proposed framework. The non-padded cases provide direct evidence that the circuit is correct and that the theoretical expression for the post-selection probability is accurate. The padded cases, in turn, clarify the effect of embedding non-power-of-two instances into the nearest power-of-two dimension and delimit the scope under which comparisons of $P_{\text{succ}}$ remain mathematically meaningful.

## 6. Conclusions

In this work, we introduced the Modulated Quantum Fourier Transform (MQFT) associated with modulated circulant matrices and used it to develop two quantum algorithms for modulated circulant matrix–vector multiplication. By exploiting the structured spectral decomposition of this matrix family, we showed that the MQFT can be implemented with asymptotic gate complexity $O(\log^2 N)$ under an oracle model for the modulation parameters, matching the complexity of the standard quantum Fourier transform. Furthermore, we also presented explicit quantum circuit constructions for the MQFT, for the modulated cyclic permutation operator, and for the corresponding matrix–vector multiplication procedure. In addition, we derived an analytical expression for the post-selection success probability, which complements the circuit-level description of the algorithm.

The numerical simulations support the correctness of the proposed framework. In the non-padded cases, the output fidelities were above 0.97, and the measured post-selection probabilities were in close agreement with the theoretical values. The padded cases further illustrated the effect of embedding non-power-of-two instances into the nearest power-of-two dimension, thereby clarifying the scope under which direct comparisons of $P_{\text{succ}}$ remain meaningful.

These results show that the spectral structure of modulated circulant matrices can be exploited to design efficient quantum procedures beyond the standard circulant setting.

**Acknowledgments**. Kimy Agudelo acknowledges partial support from the Universidad Católica del Norte through the Postgraduate Maintenance Scholarship (Resolution VRA No. 380/2024), and from ANID-Chile through the FONDECYT Regular grant No. 1241855. Acknowledge Vicerrectoría de Investigación y Desarrollo Tecnológico (VRIDT) at Universidad Católica del Norte (UCN) for the scientific support provided by Núcleo de Investigación en Simetrías y la Estructura del Universo (NISEU-UCN), Resolución VRIDT N°200/2025.

**Disclosure statement**

The authors declare no conflict of interest in the preparation of the manuscript.

## References

[1] A. W. Harrow, A. Hassidim, S. Lloyd, Quantum Algorithm for Linear Systems of Equations, Phys. Rev. Lett. 103 (15) (2009) 150502. arXiv:0811.3171, doi:10.1103/physrevlett.103.150502.

[2] G. H. Golub, C. F. Van Loan, Matrix Computations, 3rd Edition, Johns Hopkins University Press, Baltimore, 1996.

[3] S. S. Zhou, J. B. Wang, Efficient quantum circuits for dense circulant and circulant-like operators, Royal Society Open Science 4 (2017) 160906. arXiv:1607.07149, doi:10.1098/rsos.160906.

[4] L.-C. Wan, C.-H. Yu, S.-J. Pan, F. Gao, Q.-Y. Wen, S.-J. Qin, Asymptotic quantum algorithm for the Toeplitz systems, Phys. Rev. A 97 (6) (2018) 062322. arXiv:1608.02184, doi:10.1103/PhysRevA.97.062322.

[5] A. Daskin, Quantum implementation of circulant matrices and its use in quantum string processing, arXiv preprint arXiv:2206.09364 (2022). arXiv:2206.09364.

[6] E. Andrade, C. Kızılateş, C. Manzaneda, et al., On properties of $n$-parametric and bi-geometric circulant matrices, Comput. Appl. Math. 45 (2026) 185. doi:10.1007/s40314-025-03511-5.

[7] L. Hou, Z. Huang, C. Lv, Quantum Algorithms for the Multiplication of Circulant Matrices and Vectors, Information 15 (8) (2024) 453. doi:10.3390/info15080453.

[8] G. Brassard, P. Hoyer, M. Mosca, A. Tapp, Quantum amplitude amplification and estimation, Contemporary Mathematics 305 (2000) 53–74. arXiv:quant-ph/0005055, doi:10.1090/conm/305/05215.

[9] S. A. Cuccaro, T. G. Draper, S. A. Kutin, D. P. Moulton, A new quantum ripple-carry addition circuit, arXiv preprint arXiv:quant-ph/0410184 (2004). arXiv:quant-ph/0410184.

[10] J. Welch, D. Greenbaum, S. Mostame, A. Aspuru-Guzik, Efficient quantum circuits for diagonal unitaries without ancillas, New J. Phys. 16 (3) (2014) 033040. arXiv:1306.3991, doi:10.1088/1367-2630/16/3/033040.

[11] J. Martyn, Y. Liu, A. W. Chin, A. Aspuru-Guzik, Efficient fully-coherent quantum signal processing algorithms for real-time dynamics simulation, arXiv preprint arXiv:1905.05378 (2019). arXiv:1905.05378.